\documentclass[checkin,showpacs,psfig,aps,prb]{revtex4}

\newcommand{\ba}{\begin{eqnarray}}
\newcommand{\ea}{\end{eqnarray}}

\usepackage{graphicx,color}

\newcommand{\ce}{$\rm C_{60}$}
\newcommand{\cm}{$\rm C_{60}$ }

\newcommand{\agl}{$\rm A_g(1)$~}

\newcommand{\be}{\begin{equation}}
\newcommand{\ee}{\end{equation}}
\newcommand{\la}{\langle}

\newcommand{\ra}{\rangle}
\newcommand{\et}{{\it et al. }}
\newcommand{\ete}{{\it et al.}}

\newcommand{\bn}{\begin{enumerate}}
\newcommand{\en}{\end{enumerate}}

\begin{document}

\title{Optically- and thermally-driven huge lattice orbital and
  spin angular momenta from spinning fullerenes}

\author{G. P. Zhang$^*$} \affiliation{Department of Physics, Indiana State
 University, Terre Haute, Indiana 47809, USA}

\author{Y. H. Bai}
\affiliation{Office of Information Technology, Indiana State
  University, Terre Haute, Indiana 47809, USA}

\author{Thomas F. George}

\affiliation{Departments of Chemistry \& Biochemistry and Physics \&
  Astronomy, University of Missouri-St. Louis, St.  Louis, MO 63121,
  USA }

\date{\today}

\begin{abstract}
Lattice vibration in solids may carry angular momentum. But
unlike {the intrinsic spin of electrons}, the lattice vibration
is rarely rotational. To induce angular momentum, one needs to find a
material that can accommodate a twisted normal mode, two orthogonal
modes or excitation of magnons. If excitation is too strong, one may
exceed the Lindemann limit, so the material melts. Therefore these
methods are not ideal. Here, we theoretically propose a new route to
phonon angular momentum in a molecular crystal \ce. We find that a
single laser pulse is able to inject a significant amount of angular
momentum to \ce, and the momentum transfer is
helicity-dependent. Changing from right-circularly polarized light to
left-circularly polarized light switches the direction of phonon
angular momentum. On the ultrafast time scale, the orbital angular
momentum change closely resembles the displacive excitation of
coherent phonons, with a cosine-function dependence on time, different
from the spin counterpart. Atomic displacements, even under strong
laser excitation, remain far below the Lindemann criterion.  Under
thermal excitation, spinning \cm even at room temperature generates a
huge angular momentum close to several hundred $\hbar$.  Our finding
opens the door to a large group of fullerenes, from \ce, C$_{70}$ to
their endohedral derivatives, where angular momentum can be generated
through light or temperature. This paves the way to the phononic
control electronic spin and harvesting thermal energy through phonon
angular momentum.
\end{abstract}


\maketitle

Nuclear vibration is ubiquitous and important to biological, chemical,
and physical processes, ranging from photosynthesis
\cite{brixner2005}, photoisomerization \cite{gai1998} to spin
manipulation \cite{garanin2015,fahnle2017,zahn2020}.  Microscopically,
each material has its own vibrational normal modes.  However, due to
the symmetry constraint \cite{coh2019}, normal modes rarely have a
rotational eigenvector, so the phonon angular momentum (PAM) is
feeble.  The concept of phonon spin is not new
\cite{vonsovskii1962,levine1962}, and reemerged \cite{zhang2014}.
McLellan \cite{mclellan1988} investigated PAM in a rotationally
invariant harmonic model. Interest in PAM is more recent
\cite{zhang2014,nakane2018,ruckriegel2020,park2020,streib2021}.  It
was shown that one may use an electric field \cite{moseni2021} or
temperature gradient \cite{hamada2018} to control PAM. By coupling PAM
to the electronic spin degrees of freedom, one can use PAM to control
spin devices \cite{stupakiewicz2021}.  However, the main obstacle is
how to generate PAM in the first place.  It was predicted that one
could introduce chiral phonons in monolayer molybdenum disulfide
\cite{zhang2015}, but the real calculation \cite{juraschek2019} showed
that the effect is tiny in transition metal dichalcogenides.  The same
thing is true for Pt$_3$ and Pt$_5$ \cite{bistoni2021}, where the most
of vibrational modes has nearly zero angular momentum, with the
maximum close to $0.094(\hbar/2)$.  Experimentally, it was
demonstrated that one could construct a superimposed vibrational state
which consists of two orthogonal vibrational modes to generate an
artificial circular phonon mode \cite{nova2017}. One may also couple
the phonon to magnons in a magnetic material
\cite{holanda2018,streib2018,an2020,ruckriegel2020a} to induce spin
angular momentum of phonons, but one must first have magnons. In
general, these methods are not easy.

Molecular crystals are different. The weak bonding between neighboring
sites render atom motifs rotating freely. Buckministerfullerene \cm
\cite{kroto1985} is a prime example.  At room temperature, \cm is
spinning rapidly \cite{johnson1992,dresselhaus,bubenchikov2019} around
its equilibrium lattice positions.  Spinning is temperature-dependent.
As temperature cools down below 261 K, the rotation slows down, where
$^{13}$C NMR spectra of solid \cm show a significant broadening at a
chemical shift of 143 ppm \cite{tycko1991,tycko1991a,johnson1992}.
Its rotation can also be manipulated through substrates
\cite{wang2004,bozhko2011}, the tip of scanning tunneling microscope
\cite{neel2008} and encapsulation inside carbon nanotubes
\cite{zou2009}.  \cm rotation also affects its energy spectrum
\cite{lima2014} and vibrational spectra \cite{saito1994}. \cm and
C$_{70}$ are not the only fullerenes that spin.  Endohedral fullerenes
\cite{bethune1993}, with additional atoms encapsulated inside
fullerenes, offer additional kinds of rotational dynamics
\cite{hernandez1997}, with applications to organic photovoltaic
devices \cite{ross2009} and quantum computing \cite{twamley2003}.
Recently, possible light-induced superconductivity was reported in
K$_3$C$_{60}$ \cite{mitrano2016}. It is conceivable that light can
inject angular momentum into K$_3$C$_{60}$.  These readily available
molecular crystals represent a new frontier for phonon angular
momentum investigation.

In this Letter, instead of relying on artificial rotational motion of
atoms to search for phonon angular momentum, we start with a material
that rotates in the first place at room temperature. Specifically, we
employ \cm as an example, and we show that a laser pulse can induce a
significant orbital and spin angular momentum change. PAM depends on
the laser helicity. Circularly polarized light injects more angular
momentum to the system than linearly polarized light. Changing from
right- to left-circularly polarized light switches angular momentum
direction. The total angular momentum exactly follows the energy
change, thus reaffirming that it is physical. The orbital and spin
angular momenta depend on time differently. The orbital has a cosine
dependence, very similar to displacive excitation of coherent phonons
\cite{zeiger1992}. The lattice distortion of \cm is small even under
strong laser excitation, with the atom displacement far below the
Lindemann criterion \cite{lindemann1910}. The global rotation of \cm
generates several hundred of $\hbar$. Our finding here represents a
different direction for phonon angular momentum research, and is
expected to motivate experimental and theoretical investigations.

We define the total angular momentum of \cm as \be {\bf J}=\sum_i {\bf
  j}_i(t)=\sum_i {\bf r}_i(t)\times {\bf p}_i(t), \ee where ${\bf
  p}_i(t)$ is its momentum of atom $i$ at time $t$ and ${\bf r}_i(t)$
is its position vector.  The reason why we use {\bf J} instead of {\bf
  L} will become clear below. Although we do not quantize the atomic
vibration formally here, we still use phonons below.  To find the
initial coordinates $\{{\bf r}_i (0) \}$ of carbon atoms in \ce, we
use the spiral generation method proposed by Fowler and Manolopoulos
\cite{fowler} for the topological problem of fullerenes.  Here, in
brief, we first construct the adjacency matrix, and diagonalize
it. The three lowest eigenvectors correspond to the $x$, $y$ and $z$
coordinates of the carbon atoms, and then we properly rescale them
according to the radius of \ce.

\begin{figure}
\includegraphics[angle=0,width=0.8\columnwidth]{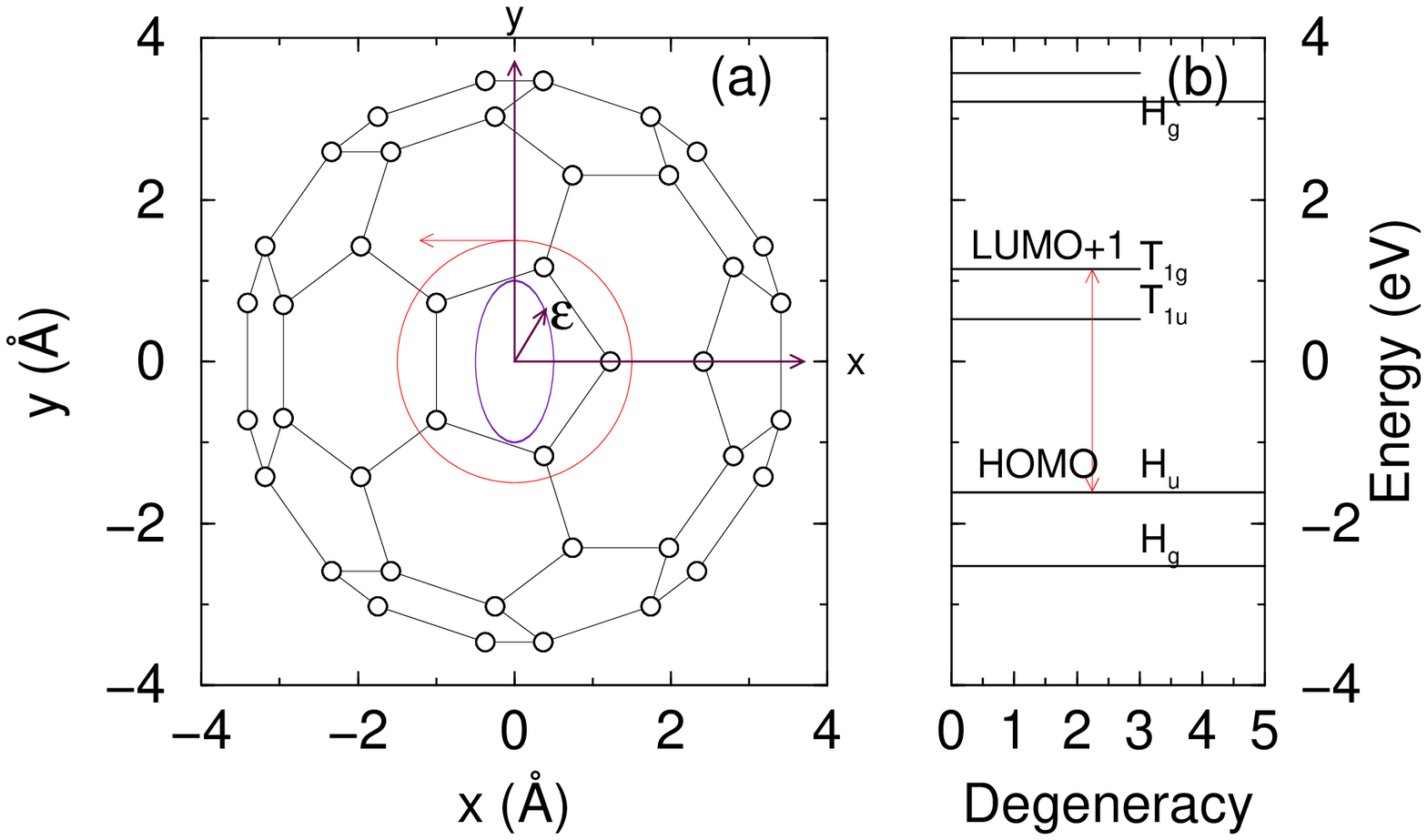}
\caption{ (a) \cm structure. The empty circles denote sixty carbon
  atoms.  The laser polarization (the vector) is in the $xy$ plane
  (the front pentagon) and controlled by the helicity $\epsilon$.  (b)
  Energy level scheme for \ce.  The double-arrow denotes the
  transition between the second lowest unoccupied molecular orbital
  (LUMO+1) and highest occupied molecular orbital (HOMO). }
\label{fig0}
\label{fig1}
\end{figure}

To optimize the geometry, we employ the Su-Schrieffer-Heeger (SSH)
model \cite{su1979} often used in conjugated polymers, where only
$\pi$ electrons are treated quantum mechanically, and the nuclear
motions are described by the classical potentials \cite{heeger1988}
since the mass of carbon atoms is more than three orders of magnitude
larger than that of the electron.  Recently, the SSH model finds an
important application in topological insulators \cite{asboth2016}.
The Hamiltonian reads \cite{prb03,prl04,prl05} \be H_0= -\sum_{\la
  ij\ra,\sigma} t_{ij} (c^{\dagger}_{i,\sigma}c_{j,\sigma} +h.  c.)+
\frac{K_1}{2} \sum_{\la
  i,j\ra}(r_{ij}-d_0)^2+\frac{K_2}2\sum_i\delta\theta_{i,p}^2+\frac{K_3}2\sum_i(\delta\theta_{i,h,1}^2+\delta\theta_{i,h,2}^2) \label{ham},
\ee where $c^{\dagger}_{i,\sigma}$ is the electron creation operator
at site $i$ with spin $\sigma(=\uparrow\downarrow)$ \cite{prl03} and
the summation $\la ij\ra$ over $i(j)$ runs from 1 to 60 with $i\ne j$.
The first term on the right-hand side represents the electron hopping
between nearest-neighbor atoms at positions ${\bf r}_i$ and ${\bf
  r}_j$, $t_{ij}=t_0-\alpha(| {\bf r}_i-{\bf r}_j|-d_0)$, where
$r_{ij}=|{\bf r}_i-{\bf r}_j|$, $t_0$ is the average hopping constant,
and $\alpha$ is the electron-lattice coupling constant.  The last
three terms are the lattice stretching, pentagon-hexagon and
hexagon-hexagon bending energies, respectively, (see further details
in \cite{sm}).  You \et
\cite{you1993} parametrized the Hamiltonian by fitting the energy gap,
bond lengths and 174 normal mode frequencies to their respective
experimental values. They found that $t_0=1.91 $ eV, $\alpha=5.0 $
eV/$\rm \AA$, $K_1= 42 $ eV/$\rm \AA^2$, $K_2= 8 $ eV/$\rm rad^2$,
$K_3= 7 $ eV/$\rm rad^2$ and $d_0=1.5532~\rm \AA$.  These parameters
are consistent with the literature values
\cite{adams1991,menon1991,feldman1992,faulhaber1993,friedman1993,harigaya1994,wang1994,liu1994}.
With You's six parameters and with the electrons initially occupying
30 lowest energy levels, we optimize the structure.  Figure
\ref{fig1}(a) shows a two-dimensional structure of our optimized \ce,
where the $z$ axis is out of the page. The \cm molecule has the
highest $I_h$ point symmetry.  60 carbon atoms form 90 bonds, 60
single bonds around 12 pentagons, and 30 single bonds shared by 20
hexagons (see Fig. \ref{fig0}(a)).  Our double bond length is 1.403
$\rm\AA$, and the single bond length is 1.443 $\rm \AA$, both of which
indeed match the experimental values of $1.40\pm 0.015 \rm \AA$ and
$1.45\pm 0.015 \rm \AA$ \cite{hedberg1991,yannoni1991}. Figure
\ref{fig1}(b) shows the energy spectrum with the degeneracy. The
highest occupied molecular orbital (HOMO) is a $H_u$ state, while the
lowest unoccupied molecular orbital (LUMO) is a $T_{1u}$ state. The
first dipole-allowed transition is between HOMO and LUMO+1, which is
highlighted by the double arrow. All these features are fully
consistent with the prior study \cite{you1993}.  The challenge is to
fit 174 normal mode frequencies \cite{menendez2000} with only six
parameters.  We find that the calculated frequencies within our theory
slightly deviate from the experimental values \cite{menendez2000}
within 3\%. For instance, our breathing mode $A_g(1)$ frequency is 481
$\rm cm^{-1}$, while the experimental one is between 492
\cite{dresselhaus} and 496 \cite{menendez2000} $\rm cm^{-1}$. Our
pentagonal pinch mode $A_g(2)$'s frequency is 1477 $\rm cm^{-1}$,
while the experimental one is 1470 \cite{dresselhaus,menendez2000}
$\rm cm^{-1}$. Here the error is 0.5 \%. Because our Hamiltonian does
not include high-order terms \cite{wu1987}, these small differences
are expected and do not affect our main conclusion of the paper. For
this reason, we do not adjust those parameters.

As seen above, there are varieties of ways to induce angular momentum
change.  We perturb the system with an ultrafast laser pulse, whose
interaction with \cm is described by \be H_I=-e\sum_{i\sigma}{\bf
  E}(t)\cdot {\bf r}_i~ n_{i\sigma}~, \label{laser}\ee where ${\bf
  E}(t)$ is the electric field of the laser and $n_{i\sigma}$ is the
electron number operator at site $i$. We choose a Gaussian pulse for
${\bf E}(t)$: ${\bf E}(t) =A_0\exp(-t^2/\tau^2)[\cos(\omega t)
  \hat{x}+\epsilon \sin(\omega t) \hat{y}]/\sqrt{1+\epsilon^2}$, where
$A_0$, $\omega$, $\tau$, $t$, $\epsilon$ and $\tau$ are the field
amplitude, laser frequency, pulse duration or width, time, helicity
and pulse duration, respectively.  $\hat{x}$ and $\hat{y}$ are the
unit vectors along the $x$ and $y$ axes, respectively. We ensure
  that $A_0$ is large enough, so the quantum effect of light is small
  and the classical treatment of our laser field is adequate. We
treat the electrons quantum mechanically through the time-dependent
Liouville equation \cite{prl03,prl05}, \be i\hbar \frac{\partial \la
  \rho_{ij}^\sigma\ra}{\partial
  t}=\langle[H,\rho_{ij}^\sigma]\rangle,\ee where $H=H_0+H_I$, and
$\rho_{ij}^\sigma $ is the density matrix. The lattice vibrations are
described by the Newtonian equation. This classical treatment of
nuclear motion is reasonable if the de Broglie wavelength is smaller
than the system size, but if the de Broglie wavelength becomes
comparable to the system size, then our approach is less accurate.
We solve the coupled Liouville and
Newtonian equations numerically.

\begin{figure}
\includegraphics[angle=0,width=0.8\columnwidth]{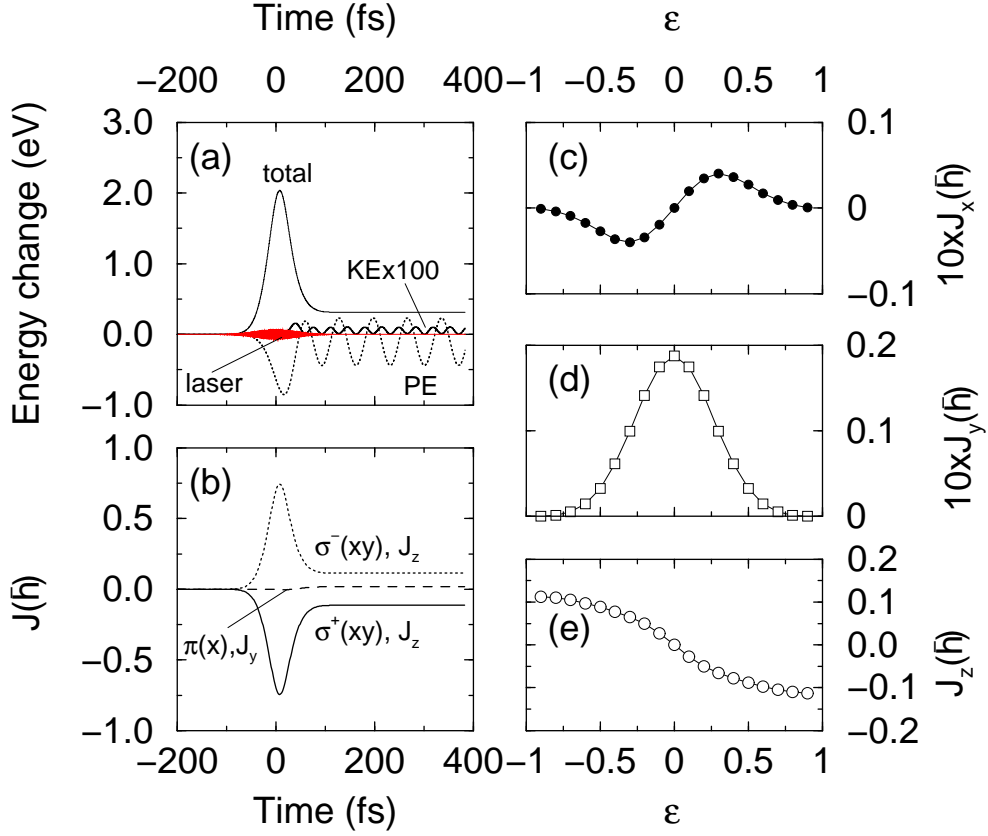}
\caption{ (a) Total energy change as a function of time under a 60-fs
  $\sigma^+$ pulse. Here $\epsilon=1$.  The laser field amplitude is
  0.01 $\rm V/\AA$, and the photon energy is $\hbar\omega=2.76$
  eV. Thin solid, thick solid, and dashed lines denote the system
  total energy, lattice kinetic and potential energy,
  respectively. The lattice kinetic energy is multiplied by 100 to
  have an easy view.  To demonstrate the energy conservation, the
  system is not damped. The red thin line around 0 is the laser
  electric field. The oscillation period in the lattice potential is
  69.3 fs, which exactly matches the frequency of the Raman-active
  $A_g(1)$ mode. (b) Total angular momentum {\bf J} under $\sigma^+$
  (solid line), $\sigma^-$ (dotted line) and $\pi$ pulses (dashed
  line).  Only the strongest component of {\bf J} is shown.  Linearly
  polarized light transfers a small angular momentum. The left- and
  right-circularly polarized light transfer opposite angular
  momenta. All the pulses have the same laser parameters as (a).
(c), (d) and (e) show the final $x$, $y$ and $z$ components of {\bf J} as a
  function of ellipticity $\epsilon$.
 }
\label{fig2}
\end{figure}

We excite \cm with a 60-fs laser pulse with electric field amplitude
of 0.01 $\rm V/\AA$. The photon energy is 2.76 eV, which is tuned to
be resonant with the transition between HOMO and LUMO+1 (see
Fig. \ref{fig1}b).  Figure \ref{fig2}(a) displays the total energy
(thin line), lattice kinetic energy (thick line) and lattice potential
energy (thick dashed line) as a function of time. The curve around
zero denotes the laser pulse.  Upon laser excitation, energy enters
the electronic system first. The peak in the total energy is mainly
due to the electron energy. Through the electron-lattice interaction
and laser excitation, the lattice starts to vibrate. Most of the
lattice energy is in the potential energy, and the kinetic energy is
very small. In the figure, we multiply the kinetic energy by
100. Carbon atoms move at a speed of $10^{-3}-10^{-2}\rm \AA/fs$, or
100-1000 m/s, very typical for nuclear vibrations.  {The
  oscillation is due to the energy exchange between the electron and
  lattice subsystems (for details, see \cite{sm}).}  Since we want to
check the energy conservation as well as vibrational oscillations, we
do not dampen our system. We find that the oscillation has a period of
69.3 fs, which exactly matches the frequency of the breathing mode
$A_g(1)$ \cite{dexheimer1993,prb06}. Such a coherent phonon excitation
has been observed experimentally \cite{dexheimer1993} and in other
systems as well \cite{zeiger1992}.  {The reason why this \agl is
  excited strongly is due to our chosen laser parameters
  \cite{prl04,sm}}.  Our interest is in the lattice angular momentum.
Figure \ref{fig2}(b) shows the lattice angular momentum for three
laser helicities ($\epsilon$): right ($\sigma^+$), left ($\sigma^-$)
and linear ($\pi$) pulses, with the polarization in the $xy$ plane
(see Fig. \ref{fig1}(a)). There is a general trend as to how the
angular momentum is transferred to the lattice. Upon laser excitation,
the flow of angular momentum to the lattice occurs on the 100 fs time
scale, which corresponds to the total energy absorbed. Although each
atom still vibrates and exchanges its angular momentum with the rest
of the atoms, the total angular momentum must remain constant and is
conserved in the absence of an external field. This is rigorously
reproduced in our calculation.  The angular momentum shows a strong
dependence on $\epsilon$. For $\sigma^-$, the lattice angular momentum
is along the $+z$ axis, while for $\sigma^+$, it is along the $-z$
axis. The $\pi$ pulse induces a much weaker angular momentum (see the
long-dashed line). This is expected because the angular momentum of
light is directly related to the helicity of light. When it interacts
with \ce, the lattice angular momentum has the hallmark of incident
light. Figures \ref{fig2}(c), (d) and (e) show the detailed change in
$J_{x}$, $J_{y}$ and $J_{z}$ as a function of helicity $\epsilon$.
When $\epsilon=0$, the laser polarization is linear along the $x$
axis. One can see that only $J_y$ differs from zero, because the atom
moves mainly along the $x$ axis. Going from $\epsilon=0$ to 1, our
pulse changes from a linearly polarized pulse to elliptically, and
finally to a circularly polarized pulse. Both $J_x$ and $J_z$ increase
significantly as the atom also moves along the $y$ axis.  $J_z$ is
negative as expected. When we change $\epsilon$ from 0 to -1, we have
a left elliptically to circularly polarized pulse, so $J_z$ flips its
sign, fully consistent with our finding in Fig. \ref{fig2}(b).

\begin{figure}
\includegraphics[angle=0,width=0.8\columnwidth]{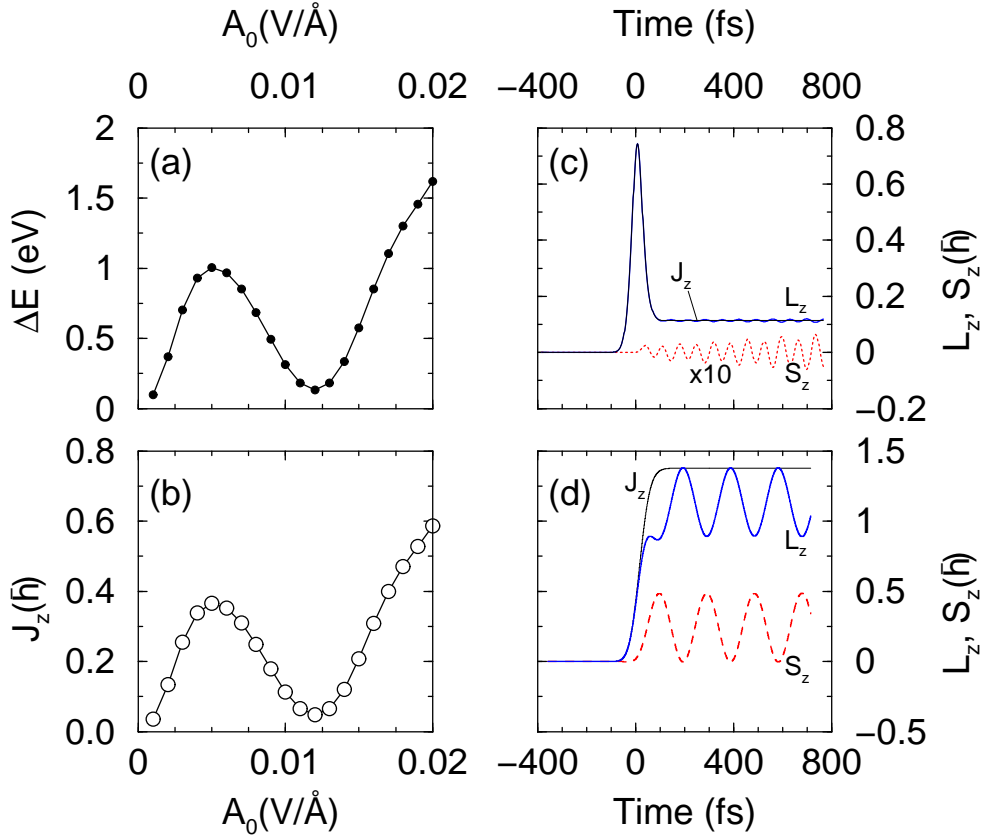}
\caption{ (a) Absorbed energy $\Delta E$ as a function of laser field
  amplitude $A_0$. The photon energy and pulse duration are the same
  as Fig. \ref{fig2}. (b) Total angular momentum as a function of
  laser field amplitude $A_0$, which matches the total energy
  change. This demonstrates that {\bf J} is physical. (c) The orbital
  $L_z$ and spin $S_z$ angular momenta as a function of time. $S_z$ is
  very small when $\hbar\omega=2.76$ eV. In the figure, it is
  multiplied by 10. (d) The orbital $L_z$ and spin $S_z$ angular
  momenta as a function of time with $\hbar\omega=0.185$ eV.  $S_z$
  reaches a comparable magnitude as $L_z$.  The dependence of $S_z$ on
  time follows a sine function, while that of $L_z$ follows a cosine
  function, very similar to DECP \cite{zeiger1992}.  }
\label{fig3}
\end{figure}

 What is unknown or less familiar is whether and how the lattice
 angular momentum {\bf J} can characterize a system.  This is an
 important conceptual question because angular momentum is rarely used
 to characterize the dynamics of a system, except in atoms and some
 simple structures. Prior studies
 \cite{zhang2014,nakane2018,ruckriegel2020,park2020,streib2021} are
 all based on this assumption. We compare {\bf J} against the absorbed
 energy $\Delta E$ into \ce.  When the laser photon energy is off
 resonance, the energy change follows the perturbative path and has a
 simple dependence on the laser field amplitude $A_0$ \cite{shen}, so
 the agreement between the angular momentum ${\bf J}$ and $\Delta E$
 could be coincidental.  The harder question is what happens if the
 laser is on resonance.  Figure \ref{fig3}(a) shows the energy
 absorbed as a function of laser field amplitude where the photon
 energy is tuned to the dipole-allowed transition between the HOMO and
 LUMO+1 states (Fig. \ref{fig1}(b)). It is clear that the dependence
 on $A_0$ is highly nonlinear. When $A_0$ is far below 0.005 $\rm
 V/\AA$, the change is linear, but as far as it closes to 0.005 $\rm
 V/\AA$, it peaks, after which it starts to decrease.  Because
 resonant excitation activates multiple real electronic excitation,
 this opens other channels and $\Delta E$ decreases. Once $A_0$
 exceeds 0.01 $\rm V/\AA$, $\Delta E$ increases again, which is
 consistent with our prior study \cite{prb03}.  Figure \ref{fig3}(b)
 is our angular momentum.  One can see that its change with $A_0$
 matches the absorbed energy change, down to some minor details. This
 proves that {\bf J} is physical, and it represents the acquired
 angular momentum from light.

It has been proposed that if one rewrites ${\bf r} (t)$ as a sum of
the equilibrium position ${\bf r}_i(0)$ and the displacement ${\bf
  u}_i(t)$, ${\bf r}_i(t)={\bf r}_i(0)+{\bf u}_i(t)$, the angular
momentum for atom $i$ can be separated into two terms: \be {\bf
  j}_i(t)={\bf r}_i(0)\times {\bf p}_i(t) +{\bf u}_i(t)\times {\bf
  p}_i(t) \equiv {\bf l}_i(t) +{\bf s}_i(t) . \label{spin} \ee ${\bf
  l}_i(t)$ is called the orbital angular momentum of atom $i$, while
${\bf s}_i(t)$ is called the spin angular momentum
\cite{mclellan1988,zhang2014,garanin2015,ruckriegel2020} because ${\bf
  s}_i$ is independent of the initial position of an atom. By separating
orbital from spin angular momentum, we can directly investigate the
interplay between spin and orbital angular momenta on an ultrafast
time scale.

Figure \ref{fig3}(c) compares the spin angular momentum with the
orbital counterpart. We employ a 60-fs laser pulse with
$\hbar\omega=2.76$ eV and $A_0=0.01\rm V/\AA$.  We notice that the
angular moment mainly enters the orbital part. From Eq. \ref{spin}, we
see that it is the momentum ${\bf p}_i$ that is mainly responsible for
this increase. $S_z$ is very small and magnified by 10 in the
figure. What is interesting is that the spin angular momentum does not
follow the laser pulse, very different from the orbital angular
momentum. The reason for this difference is straightforward. $S_z$
depends on both ${\bf u}_i(t)$ and ${\bf p}_i$. Although the pulse
  peaks at 0 fs, the atom cannot follow instantaneously and needs time
  to respond.  The behavior of $L_z$ is like a cosine function, with
  its maximum at 0 fs, similar to displacive excitation of coherent
  phonons (DECP) \cite{zeiger1992}. But $S_z$ is like a sine
  function. The sum of $S_z$ and $L_z$ is $J_z$, which is constant
  after the laser field is gone because the system does not have a way
  to exchange angular momentum.  This demonstrates that our definition
  of angular momentum through Eq. \ref{spin} is physically sound.

It has been a big challenge to induce phonon spin angular momentum in
real materials \cite{nova2017,holanda2018,juraschek2019}.  We can
increase spin angular momentum by tuning the laser photon energy to
vibrations directly \cite{stupakiewicz2021}. We choose
$\hbar\omega=0.185$ eV, or 44.7 THz, to be resonant with nuclear
vibrations.  The laser pulse duration is also 60 fs and the field
amplitude is increased to 0.15 $\rm V/\AA$. {We caution that our
  current treatment of nuclear vibration may not be adequate to
  compute the infrared spectrum since we do not include the nuclear dipole
  moment and the nuclear vibration is not treated quantum
  mechanically.} Figure \ref{fig3}(d) shows that $S_z$ reaches
0.5$\hbar$. Under our laser field, the structure of \cm is well
maintained, far below the Lindemann criterion \cite{lindemann1910} for
melting. This criterion has been a major obstacle for a prior study
\cite{juraschek2019}, where the atomic displacements for BaO and
LiNbO$_3$ can be as large as 0.1 $\rm \AA$. Therefore, an experimental
test is difficult.

\begin{figure}
\includegraphics[angle=0,width=0.4\columnwidth]{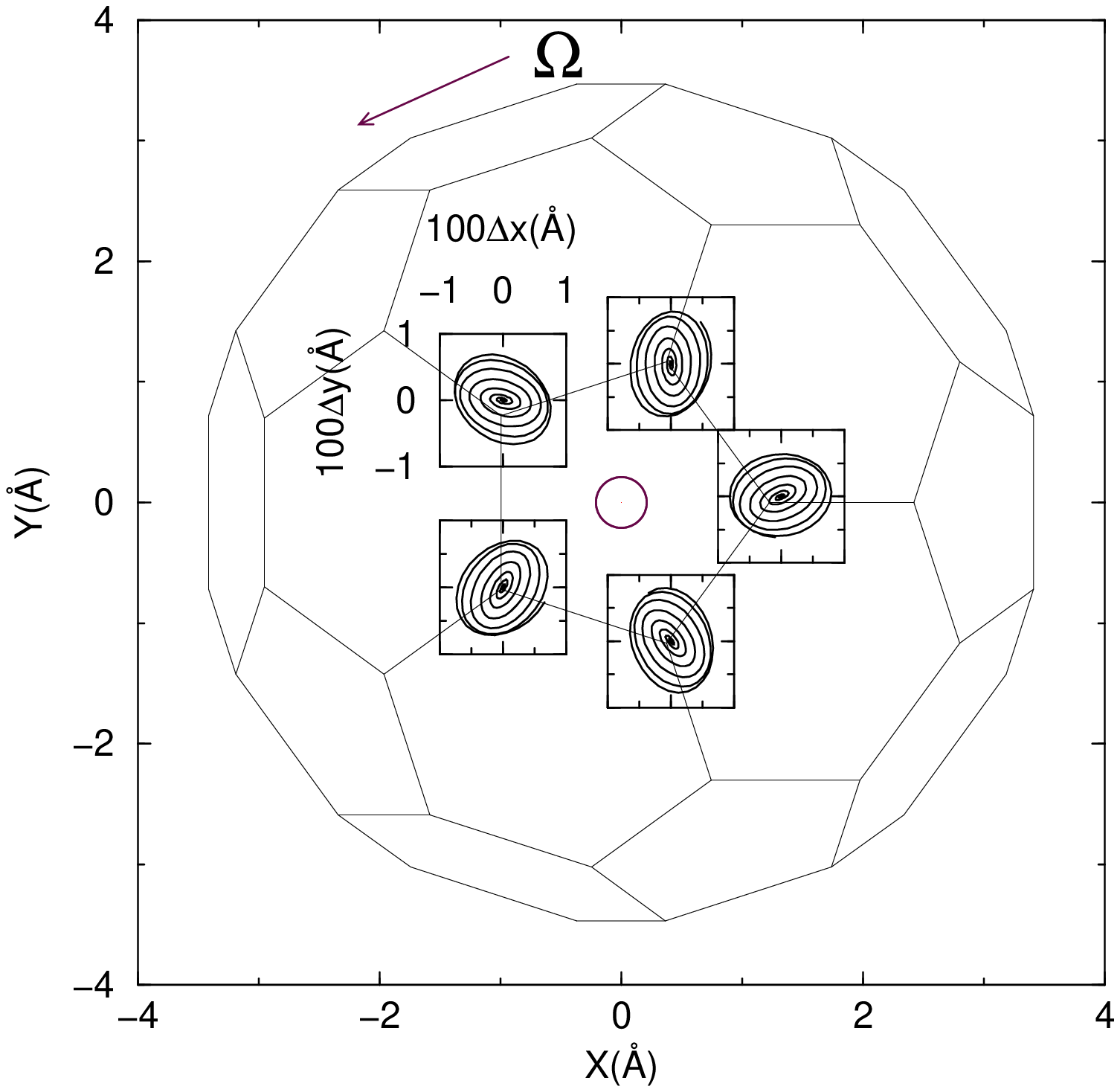}

\vspace{1cm}

\includegraphics[angle=0,width=0.6\columnwidth]{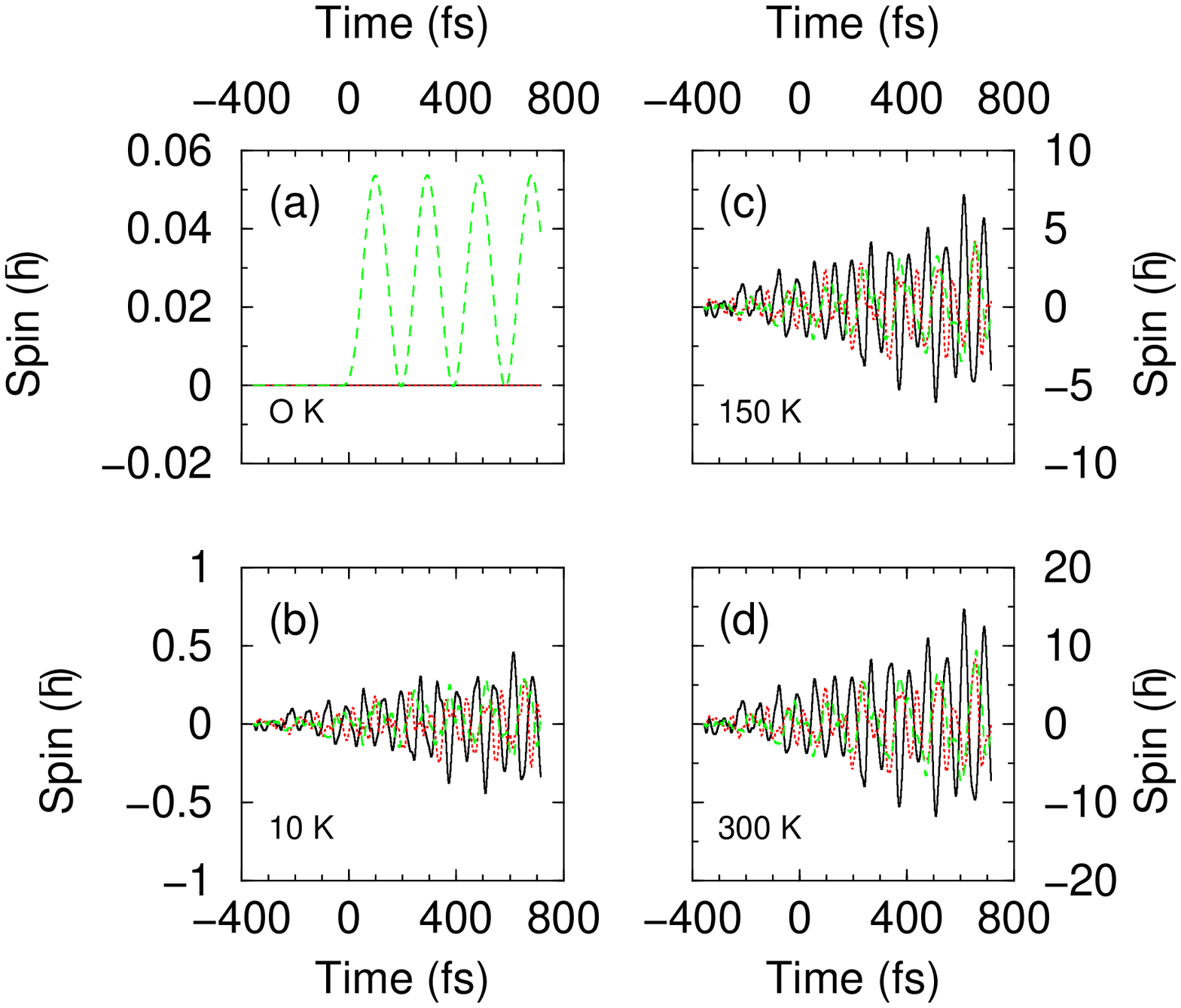}
\caption{(Top) Traces of displacements of five carbon atoms on the
  front pentagon in the earlier stage of laser excitation. The change
  is only 0.01 $\rm \AA$, which is well below the Lindemann criterion
  \cite{lindemann1910}. The arrow on the top left denotes the rotation
  of entire \ce. (Bottom) Phonon spin angular momentum at (a) 0
    K, (b) 10 K, (c) 150 K and (d) 300 K, where the solid, dotted and
    long-dashed lines denote $S_x$, $S_y$ and $S_z$, respectively.
    Its magnitude becomes larger as temperature increases. }
\label{fig4}
\end{figure}

The top figure of Fig.  \ref{fig4} shows atomic displacements
projected on the $xy$ plane for five atoms on the front pentagon. The
scales for the $x$ and $y$ axes are increased by 100 for an easy
view. One can see the displacement is well below the Lindemann
criterion. This paves the way to experimental testing. What is even
more interesting is that besides these internal rotations, under
thermal excitation, the entire molecule spins rapidly, although along
arbitrary axes at room temperature.  The angular momentum is
$L=I\Omega$, where $I$ is the moment of inertia of \cm and $\Omega$ is
the angular speed. Assuming \cm spins along the $z$ axis through two
pentagons (see Fig. \ref{fig0}), the moment of inertia is
$I_z=m\sum_{i=1}^{60} (x_i^2+y_i^2)$, where $m$ is the mass of carbon
atom and $x_i$ and $y_i$ are the coordinates of atom $i$. Note that
the moment of inertia is almost identical along other axes because \cm
is highly spherical.  An early estimate puts $\Omega$ between 1.5/ps
and 2.8/ps \cite{zhang1991}.  According to the latest theoretical
estimate \cite{bubenchikov2019}, its average angular frequency is 0.34
rad/ps. {Since our present study only contains one single \cm and
  has no intermolecular interaction between several fullerenes, we are
  unable to verify their average angular frequency.}  Instead, we use
their number and find $|{\bf J}|$ for \cm to be $322 \hbar$.  This
angular momentum is from the global rotation of \ce.  {Because
  our Hamiltonian in Eq. \ref{ham} only depends on the distance
  between neighboring atoms, the rapid global rotation of C$_{60}$
  does not enter our calculation directly and only implicitly through
  the laser field polarization.  However, due to the high symmetry of
  C$_{60}$, different polarizations yield almost identical
  results. Therefore, the global rotation must be treated separately
  in our formalism. }  To see the thermal effect of the global
rotation on the spin angular momentum, we set the initial velocities
of the carbon atoms to $\sqrt{k_BT/m}$, where $k_B$ is the Boltzmann
constant and $m$ is the carbon mass. We allow a random distribution of
the initial velocity directions.  We use four temperatures, $T=0$, 10,
150 and 300 K.

The results are shown in the bottom
figures of Fig. \ref{fig4}.  Figure \ref{fig4}(a) is our result with 0
K, but with a small laser field amplitude of 0.01 $\rm V/\AA$.  The
solid, dotted and long-dashed lines denote $S_x$, $S_y$, and $S_z$,
respectively, and are the same for the rest of the figures.  We see
that only $S_z$ has a sizable value. But as we increase temperature to
10 K, we find that all the other components are increased to 0.5
$\hbar$. The amplitude is comparable among all the three
components. At 150 K, the spin increases to $5\hbar$. At room
temperature of 300 K, Fig. \ref{fig4}(d) shows the spin angular
momentum increases to $10\hbar$.  We also check whether our initial
configuration of velocity direction affects our results. Among all the
configurations we investigate, we do not see any qualitative
difference. They should be observable experimentally. Spinning \cm
  could be a natural phonon angular momentum generator. As a large
  group of other fullerenes are available and even its endohedral
  forms already exist, such as N@\cm and P@\ce, our finding opens a
  new route to phonon angular momentum.  The future study should focus
  on how to harvest such a huge momentum for technological
  applications.

In conclusion, we have proposed a route to generate phonon orbital and
spin angular momenta from \ce, without resorting to more complicated
synthetic methods
\cite{zhang2015,juraschek2019,nova2017,holanda2018}. We show that a
single laser pulse can transfer a significant amount of angular
momentum to \ce. The phonon angular momentum that \cm receives shows a
strong dependence on laser helicity. The circularly polarized light
injects more momentum than the linearly polarized light. We
demonstrate that the total angular momentum change follows the total
energy change faithfully, thus establishing that even in the time
domain PAM is a valid concept and can be used to characterize the
system property. The orbital angular momentum is normally larger than
the spin counterpart, and it also has a different dependence on
time. Similar to DECP \cite{zeiger1992}, the orbital has a cosine
function dependence, while the spin has a sine function dependence. We
find that the atomic displacement is spiral and its amplitude is far
below the Lindemann criterion, which has been a big obstacle to a
prior study \cite{juraschek2019}. The angular momentum of the entire
\cm reaches 322 $\hbar$ at room temperature. Since endohedral
fullerenes such as N@\cm and P@\cm are readily available, our study
points out a large group of materials suitable for phonon angular
momentum generation. It is expected that our finding will motivate
experimental and theoretical investigations in this field and beyond.

\acknowledgments

This work was solely supported by the U.S. Department of Energy under
Contract No. DE-FG02-06ER46304. Part of the work was done on Indiana
State University's high performance Quantum and Obsidian clusters.
The research used resources of the National Energy Research Scientific
Computing Center, which is supported by the Office of Science of the
U.S. Department of Energy under Contract No. DE-AC02-05CH11231.

$^*$ guo-ping.zhang@outlook.com. https://orcid.org/0000-0002-1792-2701

\end{document}